\documentclass[12pt]{iopart}

\usepackage[utf8]{inputenc} 

\usepackage{wasysym}
\usepackage{siunitx}

\usepackage{graphicx}

\newcommand{\mytitle}{Comments on the NEMA NU 4-2008 Standard on Performance Measurement of Small Animal Positron Emission Tomographs}
\newcommand{\myauthor}{Patrick Hallen}

\usepackage{subcaption}

\usepackage{afterpage}

\usepackage{hyperref}
\hypersetup{colorlinks,
	citecolor=black,
	linkcolor=black,
	filecolor=black,
	urlcolor=black,
	pdftitle={\mytitle},
	pdfauthor={\myauthor},
	pdfcreator={\myauthor},
	final,
}

\begin{document}

\vspace*{-2cm}
\title[Comments on NEMA NU 4-2008]{\mytitle}

\author{Patrick Hallen$^{1}$,
		David Schug$^{1,2}$,
		Volkmar Schulz$^{1,2,3,4}$}
\address{$^1$Department of Physics of Molecular Imaging Systems, Institute for Experimental Molecular Imaging, RWTH Aachen University, Aachen, Germany}
\address{$^2$Hyperion Hybrid Imaging Systems GmbH, Aachen, Germany}
\address{$^3$III. Physikalisches Institut B, RWTH Aachen University}
\address{$^4$Fraunhofer Institute for Digital Medicine MEVIS, Aachen, Germany}
\ead{patrick.hallen@pmi.rwth-aachen.de, volkmar.schulz@pmi.rwth-aachen.de}

\begin{abstract} 
	The National Electrical Manufacturers Association's (NEMA) NU 4-2008
	standard specifies methodology for evaluating the performance of
	small-animal PET scanners. The standard's goal is to enable comparison of
	different PET scanners over a wide range of technologies and geometries
	used. In this work, we discuss if the NEMA standard meets these
	goals and we point out potential flaws and improvements to the standard.

	For the evaluation of spatial resolution, the NEMA standard mandates the
	use of filtered backprojection reconstruction. This reconstruction method
	can introduce star-like artifacts for detectors with an anisotropic
	spatial resolution, usually caused by parallax error. These artifacts can
	then cause a strong dependence of the resulting spatial resolution on the
	size of the projection window in image space, whose size is not fully specified in the
	NEMA standard. If the PET ring has detectors which are perpendicular to a
	Cartesian axis, then the resolution along this axis will typically improve with
	larger projection windows.

	We show that the standard's equations for the estimation of the random rate for PET
	systems with intrinsic radioactivity are circular and not satisfiable.
	However, a modified version can still be used to determine an approximation
	of the random rates under the assumption of negligible random rates for
	small activities and a constant scatter fraction. We compare the resulting
	estimated random rates to random rates obtained using a delayed coincidence window
	and two methods based on the singles rates. While these methods give
	similar esimates, the estimation method based on the NEMA
	equations overestimates the random rates.

	In the NEMA standard's protocol for the evaluation of the sensitivity, the
	standard specifies to axially step a point source through the scanner and
	to take a different scan for each source position. Later, in the data
	analysis section, the standard does not specify clearly how he
	different scans have to be incorporated into the analysis, which can lead to unclear
	interpretations of publicizeded results.

	The standard's definition of the recovery coefficients in the image quality
	phantom includes the maximum activity in a region of interest, which causes
	a positive correlation of noise and recovery coefficients.  This leads to
	an unintended trade-off between desired uniformity, which is negatively
	correlated with variance (i.e. noise), and recovery.

	With this work, we want to start a discussion on possible
	improvements in a next version of the NEMA NU-4 standard.
\end{abstract}

\maketitle
\section{Introduction}
The National Electrical Manufacturers Association's (NEMA) NU 4-2008 standard
on ``Performance Measurements of Small Animal Positron Emission Tomography"
specifies ``standardized methodology for evaluating the performance of positron
emission tomographs (PET) designed for animal imaging"
\cite{national_electrical_manufacturers_association_nema_nema_2008}. The
standard's goal is to enable comparison of the performance of different PET
systems over a wide range of technologies and geometries used. Thus, the
methods specified in the standard should not artificially favor or disfavor
certain choices in scanner geometry and technology and the performance results
should indicate the expected performance in real-world applications as closely
as possible. Virtually all commercial small-animal PET systems and most
research prototype PET systems have published  performance evaluations based on
the NEMA standard and Goertzen et al. \cite{goertzen_nema_2012} have published a review
comparing small-animal PET systems based on the respective NEMA performance
publications. These publications are an essential benchmark in the development
of new PET systems and an important tool for the purchase decisions of potential
buyers.

The NEMA standard specifies 5 measurements with respective analysis: evaluation
of spatial resolution; evaluation of total, true, scattered, random and
noise-equivalent count rates; evaluation of system sensitivity; and quantitativ evaluation
of image quality in a standardized imaging situation using a hot-rod phantom.

The standard was devised over 10 years ago, so it does not incorporate newer
technological developments and paradigm shifts.  For instance, the use of data
acquisition into sinograms and filtered backprojection reconstruction mandated
in the standard was more
widespread than it is today. Nowadays, these methods are often only implemented
to evaluate the PET performance based on NEMA but never actually used for
real-world applications

In this work, we examine if the NEMA standard meets its goals to enable a fair
comparison of PET systems and we point out potential flaws and improvements in
the standard. In our opinion, the standard is underspecified in several parts,
limiting the comparability of different systems, since the investigators
performing the performance evaluations are still free to choose parameters
which significantly influence the results. The methods specified for evaluation
of the spatial resolution disadvantages certain system
geometries, where those geometries don't exhibit the same reduction in spatial
resolution in real-world applications. The definition of random rates is
circular and allows the use of very different other methods generating
different results. The chapter on sensitivity is
ambiguous, leading to publications using different or even unclear methods for
the measurement of sensitivity, creating ambiguity in the interpretion of sensitivity of
different PET systems.

If applicable, we demonstrate the claimed issues with simple simulation
studies.
All discussions in this work should be universally applicable to any
PET system. However, it is still helpful and instructive to support the
claims in this work with real-world data. This is done using data obtained
with the Hyperion II$^\text{D}$ PET/MRI scanner, which was developed by our
group \cite{weissler_digital_2015}. Using the same data, we
already have published a performance evaluation based on the NEMA standard
\cite{hallen_pet_2018}.

The goal of this work is to start a discussion on a revised version of the NEMA
standard and to provide input for this discussion.

\section{Spatial Resolution}

To evaluate the spatial resolution, the NEMA standard mandates the use of
point source scans which are reconstructed using filtered backprojection.
However, basically all modern PET scanners instead use an iterative maximum
likelihood expectation maximization (MLEM) algorithm for reconstruction
\cite{nagy_performance_2013, ko_evaluation_2015,
omidvari_pet_2017, spinks_quantitative_2014, miyake_performance_2014,
sato_performance_2016, hallen_pet_2018, mackewn_pet_2015,
wong_engineering_2012, prasad_nema_2011, szanda_national_2011,
krishnamoorthy_performance_2018}, so a scanner's spatial resolution using
filtered backprojection is not necessarily indicative of its spatial resolution
for applications. While the mandated filtered backprojection is intended to
benchmark the detector performance alone, we will demonstrate in the following
that it disadvantages certain scanner geometries. Furthermore, the NEMA
standard specifies that the spatial
resolution must be  determined using the projections of the reconstructed point
sources inside a window in image space, without strictly specifying the size of this projection window. We will
demonstrate that this can lead to an ambiguous spatial resolution which
depends on the size of the projection window and allows for artificially
enhancing the spatial resolution by choosing a particularly large projection
window for certain scanner geometries.

\begin{figure}[tbh]
	\centering
	\includegraphics[width=85mm]{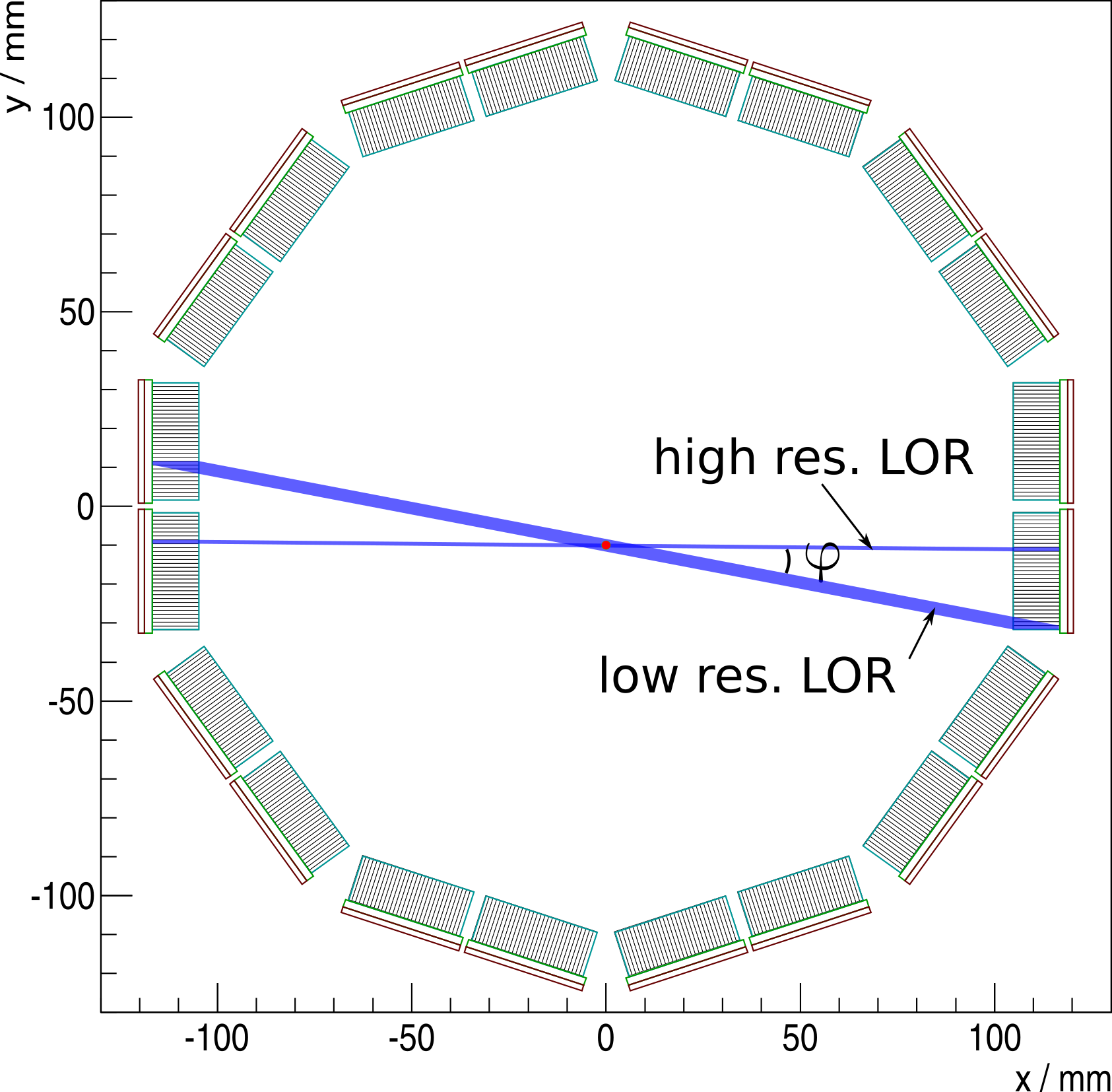}
	\caption{Ring geometry that was used for the simulations and the
	measurement. The blue bands show the parallax error of LORs, which
increases approximately proportional to the angle $\varphi$ to the normal of the block detector.}
	\label{fig:hyperion}
\end{figure}

The main disadvantage of filtered backprojection is that it doesn't include any
model of the detector and assumes an ideal, ring-like PET scanner, while
the detectors in real-world PET scanners are usually in a block geometry with anisotropic spatial
resolutions. Line of responses (LORs) perpendicular to the detectors front
face are detected with the highest resolution, while tilted LORs have a
parallax error in the detected position, which increases with more tilt of
the LORs relative to the detector's front faces as illustrated in
Figure~\ref{fig:hyperion}. In principle, this effect
can be reduced by detectors which are able to determine the depth of
interaction (DOI) of the gamma interaction, but in practice most PET system
don't employ detectors with DOI determination
\cite{nagy_performance_2013, ko_evaluation_2015, hallen_pet_2018, mackewn_pet_2015, wong_engineering_2012, szanda_national_2011, bao_performance_2009}.
Additionally, PET rings have gaps between the detector, where no LORs are detected at all.

These issues with filtered backprojection will lead to artifacts in the
reconstructed activity. For instance, each angle where the PET ring has an enhanced
spatial resolution creates an excess in reconstructed activity along the line
connecting this position with the point source and each angle with degraded
spatial resolution creates a reduction in reconstructed activity along the
respective line. Similarly, gaps between the detector create a lack of
reconstructed activity along these lines.

\afterpage{%
\begin{figure}[tbh]
	\centering
	\includegraphics[height=0.78\textheight]{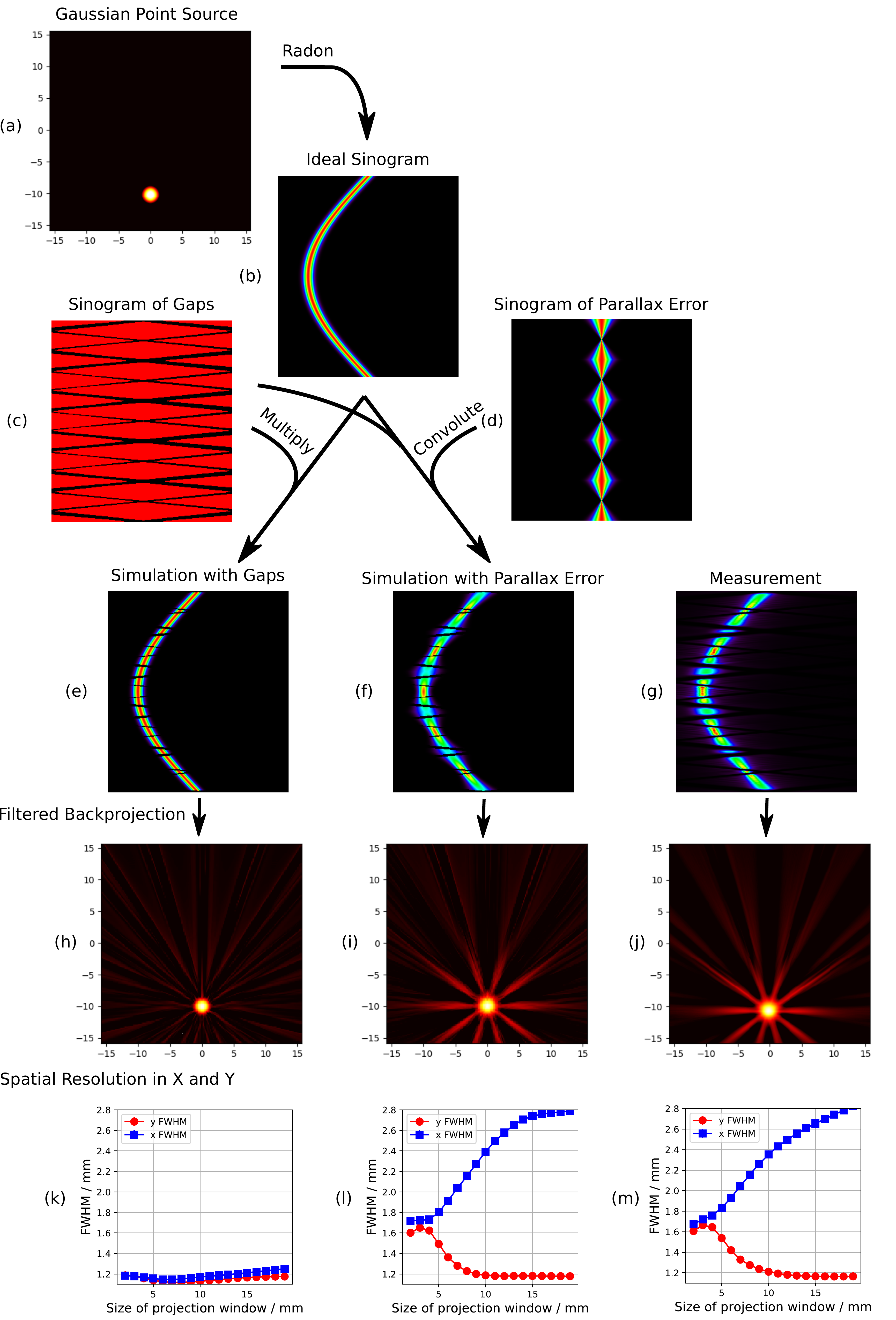}
	{%
		\phantomsubcaption\label{fig:gaussian_point_source}
		\phantomsubcaption\label{fig:ideal_sinogram}
		\phantomsubcaption\label{fig:gaps_sinogram}
		\phantomsubcaption\label{fig:psf_sinogram}
		\phantomsubcaption\label{fig:simulation_gaps}
		\phantomsubcaption\label{fig:simulation_gaps_doi}
		\phantomsubcaption\label{fig:measurement_sinogram}
		\phantomsubcaption\label{fig:simulation_gaps_FBP}
		\phantomsubcaption\label{fig:simulation_gaps_doi_FBP}
		\phantomsubcaption\label{fig:measurement_FBP}
		\phantomsubcaption\label{fig:simulation_gaps_projection}
		\phantomsubcaption\label{fig:simulation_gaps_doi_projection}
		\phantomsubcaption\label{fig:measurement_projection}%
	}
	\caption{Visualization of the influence of anisotropic detector
		resolution on the filtered backprojection and resulting spatial resolution
		along the two axis. Figure~e, h, and k show the simulation with only gaps,
		Figure~f, i, and l show
		the simulation with anisotropic detector resolution and
		gaps of 10 detector modules, and Figure~g, j, and m show a measurement.
		The simulation with
		anisotropic detector resolution and the measurement exhibit a star-like
		artifact in the reconstruction, which leads to a split in spatial resolution
	along x- and y-axis, as shown in the bottom row.}
	\label{fig:spatial_resolution}
\end{figure}
}

To understand and demonstrate this behaviour, it
is instructive to look at these effects in sinogram space. In sinogram
space, the enhanced spatial resolution of perpendicular LORs manifests as
hot-spots or rather peaks in the center of each detector modules as
Figure~\ref{fig:measurement_sinogram} shows. With increasing distance from the
center of the detector module the spatial resolution degrades, blurring the line
of the point source in sinogram space. This is equivalent to
the convolution of the sinogram of a Gaussian point source and the parallax
error of the detector. The parallax error of the detector stack
can be modeled as the shape of two triangles, connected at their tips as shown in
Figure~\ref{fig:psf_sinogram}. The parallelax error is proportional to
$\sin \varphi$, where $\varphi$ is the angle to the normal of the block
detector as defined in Figure~\ref{fig:hyperion}. The parallax error shown
in Figure~\ref{fig:psf_sinogram} is a small-angle approximation of this.

In addition to the inherent problems of mandating the use of
filtered-backprojection in the NEMA standard, the standard additionally
mandates projecting the reconstructed three-dimensional activity onto different
one-dimensional axes using a
projection window. However, the size of the projection window is not fully
specified: "The response function is formed by summing all one-dimensional
profiles that are parallel to the direction of measurement and \textit{within at
least two times} the FWHM of the orthogonal direction"
\cite[p. 7]{national_electrical_manufacturers_association_nema_nema_2008}.
The first issue is that this definition is circular,
since the minimal size of the projection window to determine the FWHM is
defined using the FWHM itself. One can easily fix this problem, either using a
sufficiently large projection window in the first place, or by reducing the
size of the projection window iteratively in dependence of the determined FWHM
in the previous iteration. However, the much bigger problem is that the size
of the projection window can strongly influence the resulting spatial
resolution. The cause of this is the integration of the star-like artifacts
created by the anisotropic spatial resolution, as we demonstrate with the
following simulation, shown in Figure~\ref{fig:spatial_resolution}.

We created the activity distribution of an ideally reconstructed point
source by assuming a rotationally symmetric two-dimensional normal
distribution, shown in Figure~\ref{fig:gaussian_point_source}.
The position of the point source is off-center at a radial offset of 10\,mm.
To investigate the essence of the effects, we don't include noise in our
simulation. From this ideally reconstructed point source we create a sinogram
by forward projection (i.e. by applying a Radon transformation). The resulting
sinogram is shown in Figure~\ref{fig:ideal_sinogram}.

We include the gaps between the detector
stacks in our simulation by creating a sensitivity sinogram, where all bins
corresponding to gaps are 0 and bins corresponding to sensitive detector
area are 1 shown in Figure~\ref{fig:gaps_sinogram}. The simulated geometry is
depicted in Figure~\ref{fig:hyperion} and follows the geometry of the
Hyperion~II$^\text{D}$ scanner to allow a comparison between simulation and
measurement.
When we include this model of gaps in our simulation by multiplying the
sensitivity sinogram with our point-source sinogram
(Figure~\ref{fig:simulation_gaps}) and then performing a filtered
backprojection (i.e. an inverse Radon transformation), we get a reconstructed
point source with slight artifacts, shown in
Figure~\ref{fig:simulation_gaps_FBP}. As stated above, the artifacts are a lack
of reconstructed activity along the lines connecting the gaps and the point
source. When analyzing the spatial resolution of the filtered backprojection
with gaps we observe little influence of the gaps compared to the filtered
backprojection of an ideal sinogram without gaps. More importantly, the
resulting spatial resolution of 1.2\,mm FWHM is stable to changes in the size of the
projection window, as shown in Figure~\ref{fig:simulation_gaps_projection}.
Thus, gaps between the detectors are not the cause of severe artifacts and only
have a very minor influence on the resulting spatial resolution with the usually
small gaps of PET scanners.

When we additionally include the effect of the anisotropic detector resolution due to
parallax errors by convolving the point-source sinogram and the point
spread function in Figure~\ref{fig:psf_sinogram}, the resulting filtered
backprojection in Figure~\ref{fig:simulation_gaps_doi_FBP} exhibits a
star-like artifact, i.e. the lines connecting the center
of each detector stack and the point source exhibit a visible excess in
activity.

\begin{figure}[tbh]
	\centering
	\includegraphics[width=85mm]{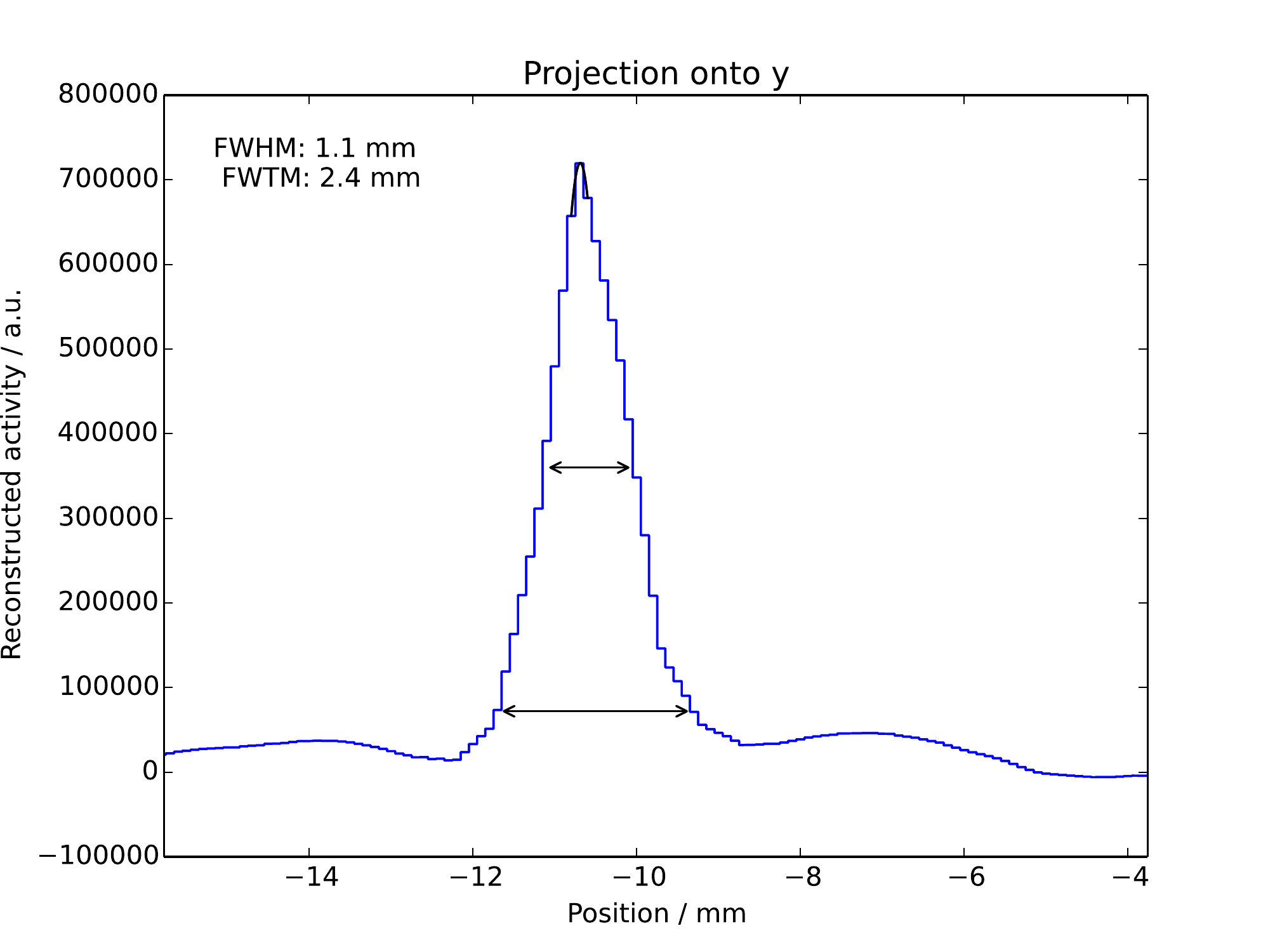}
	\caption{Line profile of the reconstructed point source projected onto the
	y-axis. The star-like artifact which is aligned with the x-axis creates an
excess in activity at the peak of the profile boosting the spatial resolution.}
	\label{fig:profile_y}
\end{figure}

If one of these excesses aligns with one of the Cartesion projection axis, and with the simulated
geometry they do so for the x-axis, the projection onto the axis perpendicular to this
axis will result in a peaked excess at the maximum of the line profile, as
shown in Figure~\ref{fig:profile_y}. A scanner's spatial resolution is
defined by the FWHM and FWTM of this profile, which depends strongly on the
height of the maximum. Therefore, a peaked excess of the maximum will
significantly enhance the resulting spatial resolution. For our geometry,
this enhancement is only observed for the y-axis, because only the x-axis
has an excess in activity aligned with it, as there aren't any detector stack
which are perpendicular to the y-axis. This difference between the
resolution in x and y is essentially an artifact and basically non-existant in
real-world applications using an iterative maximum-likelihood expectation
maximization (MLEM) reconstruction. Even worse, the extent of this effect
depends strongly on the size of the projection window as demonstrated in
Figure~\ref{fig:simulation_gaps_projection}. Increasing the size of the projection
window enhances the resulting spatial resolution in y (i.e. decreases FWHM and FWTM)
while degrading the spatial resolution in x. This makes comparison of
the spatial resolution of different PET system difficult and maybe even impossible, as
the NEMA standard does neither specify a clear projection window size nor
does it mandata that the used window size should be reported.
Thus, most publications do not state the used projection window.
\cite{nagy_performance_2013, bao_performance_2009, omidvari_pet_2017,
szanda_national_2011}.
Other geometries may not exhibit this
behavior at all, favoring or disfavoring systems which have detectors
perpendicular to a Cartesian axis. One cannot even say that systems
exhibiting this behavior have strictly worse results, as such results can
be sold in abstracts and conclusions (and marketing brochures) as "spatial
resolution up to", cherry picking the artificially
inflated spatial resolution along one of the axes.

The measurement and filtered backprojection reconstruction of point sources
with the Hyperion II$^\text{D}$ scanner shown in
Figure~\ref{fig:measurement_sinogram} and~\ref{fig:measurement_FBP} look very
similar to the simulation which includes parallax error and gaps: The sinogram
has the same hot spots at the angles where the line of responses are
perpendicular to the detector surface and the reconstruction exhibits the same
star-like artifact. The analysis of the reconstruction yields the same
observed difference in spatial resolution between the x- and y-axis.
Additionally, we observe the same strong dependence on the size of the
projection window, shown in Figure~\ref{fig:measurement_projection}.

An extreme example of a scanner geometry affected by this issue would be a box
geometry instead of the conventional ring geometry, i.e. a PET scanner with 4
large perpendicular detector modules without DOI capabilities. With such a
geometry, the filtered backprojection artifact would have the shape of a cross,
with both lines of excessive activity aligned with the x- and y-axis. Thus, the
artifact would enhance the resolution along both x- and y-axis by boosting the
maximum of both projections. 
This scenario is not solely hypothetical, as small-animal PET scanners with the
described box-like geometry exist such as PETbox 4 \cite{gu_nema_2013}. In
PETbox's NEMA NU-4 performance evaluation they state that using FBP was not
possible "since a FBP algorithm specific for the PETbox4 system with the
unconventional geometry has not been developed" \cite[p.
3797]{gu_nema_2013}.

Other examples of published performance evaluation which  have omitted the
filtered backprojection altogether when evaluating the spatial resolution are
\cite{spinks_quantitative_2014, espana_digipet:_2014}. This is an
indication that these groups don't find the results based on filtered
backprojection not indicative for the performance of their system.

Fixing the issues of this method and proposing a better method to evaluate
the spatial resolution is challenging. The NEMA standards committee surely
knew many of these issues and we believe most of the PET community will be
aware of issues with filtered backprojection, as well. However, so far, none of the performance publications based on NEMA
discussed the issues presented here, so we believe it is worthwhile to state them
to start a discussion.

One obvious solution would be to simply not use filtered backprojection and
to perform the reconstruction with a modern iterative reconstruction
instead, as all real-world measurement would be performed with this
iterative product reconstruction anyway. However, modern iterative
reconstructions include features like resolution recovery, which, in
theory, are able to reconstruct a point source as a perfect point source in
the limit of infinite statistics. Thus, the reconstruction of a point source
would mostly be a benchmark of the reconstruction and not of the underlying
detector performance. We suspect that these arguments were the main reason
why the NEMA standards committee chose filtered backprojection instead.

One alternative could be the evaluation of spatial resolution using a Derenzo
hot-rod phantom. The standard could specify the geometry of such a phantom,
specify the activity and scan time, allow the use of the reconstruction method
supplied by the manufacturer and then define a quantitative analysis method.
The Derenzo phantom is already well-established in the community as a benchmark
to evaluate the spatial resolution. For instance, several NEMA performance
publications already include such a measurement as a benchmark of spatial
resolution \cite{nagy_performance_2013, omidvari_pet_2017,
wong_engineering_2012, krishnamoorthy_performance_2018}. However, these results
are not easily comparable, as there currently isn't a standardized quantitative
analysis method to determine the spatial resolution from a measurement of a
Derenzo phantom. Usually, the spatial resolution is estimated by making a
qualitative judgement at which distance the hot rods are still discernible.
In principle, such a definition of spatial resolution based on the ability to
resolve to close points is very reasonable and
commonly used as a definition of spatial or angular resolution for telescopes
and microscopes \cite{lord_rayleigh_xxxi._1879, born_principles_1999}. However, for
a quantitative definition of spatial resolution there must be a standardized
limit of the peak-to-valley ratio between two resolvable point sources, i.e.
how much the intensity between the two peaks must dip to make them just
resolvable. In optics, there are two
commonly used criteria: The Rayleigh criterion with an intensity dip of
26.5\% and the Dawes criterion with an intensity dip of 5\%
\cite[p. 409]{mckechnie_telescope_2016}. In a standardized definition of
PET spatial resolution, the PET community could follow a similar criterion with
either the same intensity dip of 26.5\% for consistency, or standardize a new arbitrary limit.

\begin{figure}[tbh]
	\centering
	\subcaptionbox{Reconstruction of Derenzo phantom scan. The red lines show an
		example of profile lines which would be used for determination of
		valley-to-peak ratios to evaluate the spatial
		resolution.\label{fig:derenzo_profiles}}
		{\includegraphics[width=0.42\textwidth]{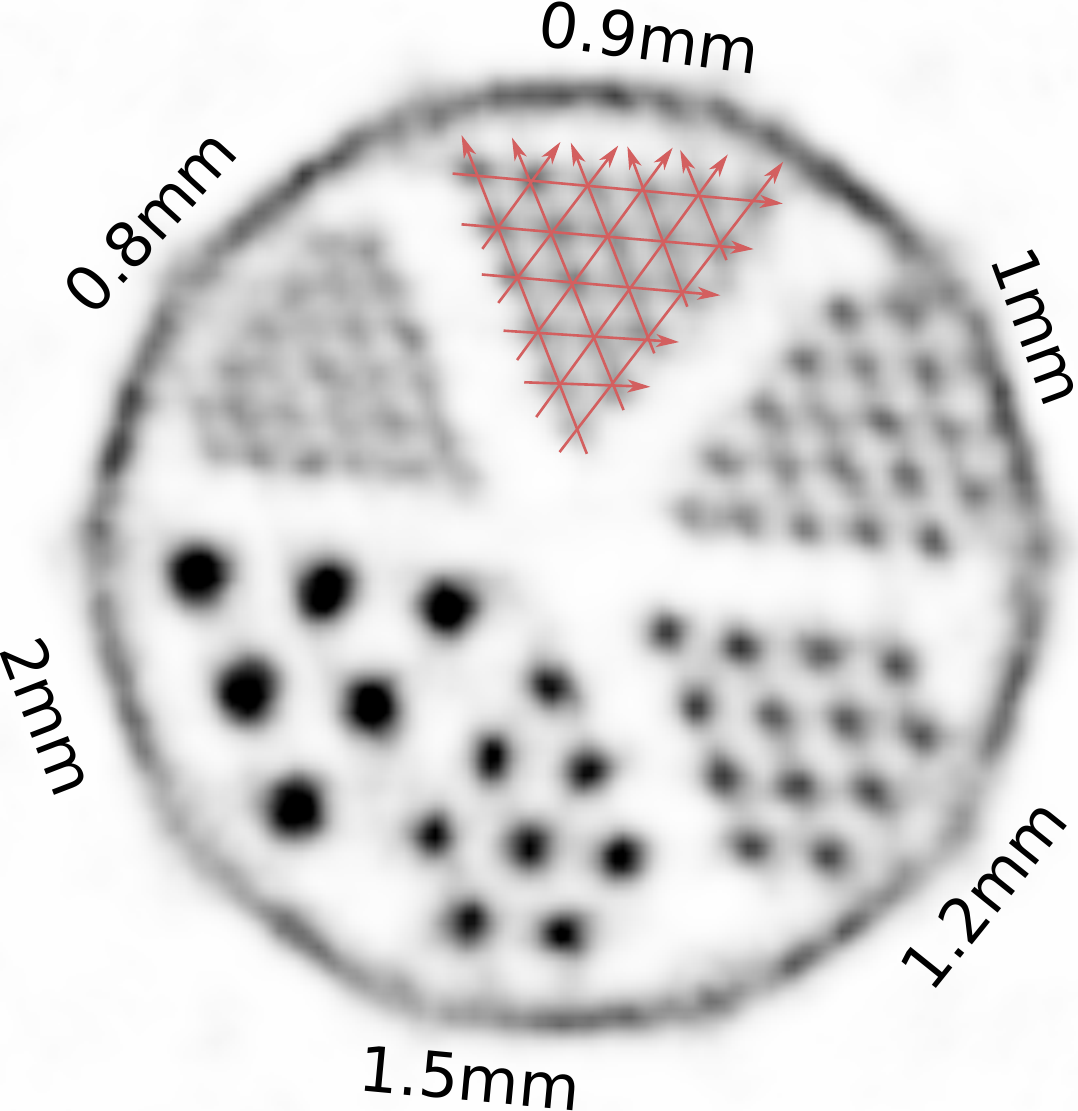}}
	\hspace{0.05\textwidth}
	\subcaptionbox{Distribution of valley-to-peak ratios for the region with a rod
	distance of 0.9\,mm. All ratios are below 0.735, which is marked with a
	red vertical line.\label{fig:derenzo_distribution}}
	{\includegraphics[width=0.5\textwidth]{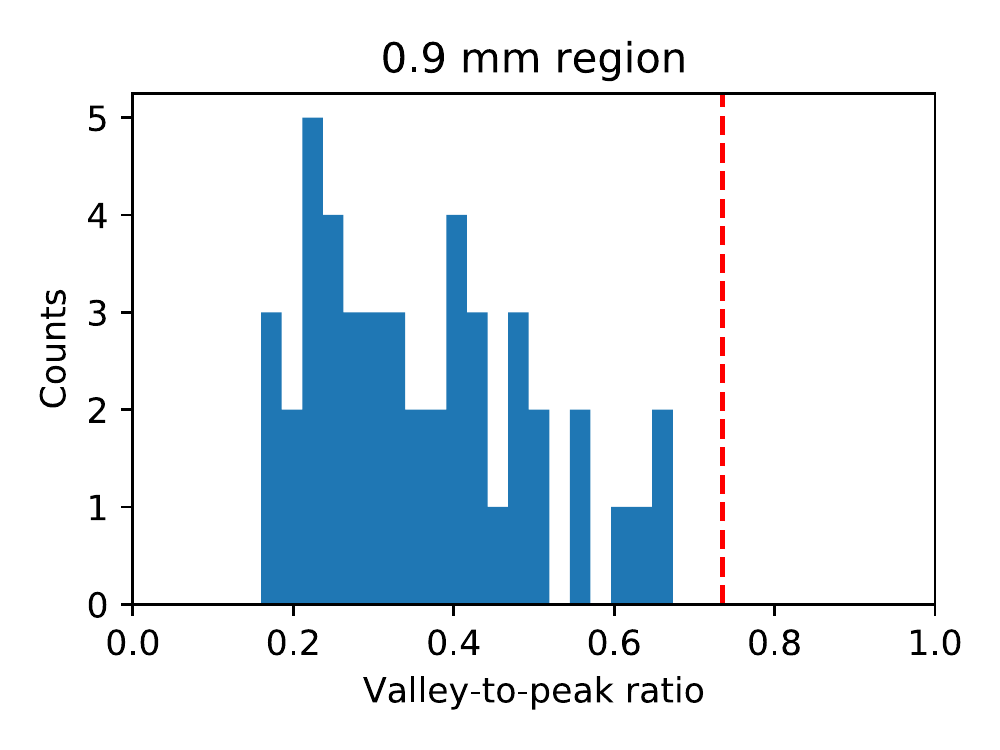}}
	\caption{Evaluation of spatial resolution using a Derenzo phantom}
\end{figure}

For the scan of a Derenzo phantom, such a resolvability criterion would require
to determine the valley-to-peak ratios of the profile lines over the different regions of
the phantom. To include anisotropies in the spatial resolution, the profile
lines should be defined over multiple angles as demonstrated in
Figure~\ref{fig:derenzo_profiles}. Figure~\ref{fig:derenzo_distribution} shows
the resulting distribution of valley-to-peak ratios for the phantom's 0.9-mm
region. We would recommend that the spatial resolution is defined as the
hot-rod distance in the region where at least 90\% of the peak-to-valley ratios are
below 0.735, i.e the valley dips are above 26.5\% for consistency with the Rayleigh criterion.
Alternatively, one could define a limit based on the average
peak-to-valley ratio of a region or using a different percentile than the
suggested 90\%. As shown in Figure~\ref{fig:derenzo_distribution}, the region
with distances of 0.9\,mm has 100\% of the valley-to-peak ratios below 0.735.
For the 0.8\,mm-region, over half of the valley-to-peak ratios would be above
0.735 in our measurement. Thus, the resulting spatial resolution would be 0.9\,mm.

To prevent arbitrary selection of peaks and valleys in a noisy
reconstruction, the standard could specify a limit for the allowed deviation
from the physical hot-rod distances when selecting the position of peak and
valleys in the profiles of the Derenzo region.

To evaluate the influence of radial and axial offsets on the spatial resolution, the
standard could specify different radial distances at which the Derenzo phantom
should be placed. Similarly, the standard could also specify additional
measurements of the rotated phantom to investigate the
isotropy of the spatial resolution.

In our opinion, such a method would depend much less on the system's
geometry and technology and would provide a much more realistic benchmark,
closely mirroring real-world use of the system. As one of the disadvantages,
the precision of this method would be limited by the differences in hot-rod
distances between the phantom's region. However, with commonly used Derenzo
phantoms one would achieve a precision in the determination of the spatial
resolution of 0.1\,mm, which is more than adequate
to assess the scanner's viability for intended applications.

As another alternative, Lodge et al. \cite{lodge_measuring_2018} have recently proposed
a novel method for the measurement of clinical PET spatial resolution using a
homogeneous cylinder phantom at an oblique angel. Such methods should also be taken into
consideration for an updated version of the standard.

\section{Scatter Fraction, Count Losses, and Random Coincidence Measurements}
The definitions of the random rate, scatter rate and scatter fraction are circular and not
satisfiable, and thus
ill-defined for systems employing detector material containing intrinsic
radioactivity, such as LYSO or LSO scintillators, as most modern PET systems
do.

To explain this issue, we give a brief summary of the NEMA standard for the
measurement of the scatter fraction, count losses and random coincidence rate
in the following. The measurement is specified as a scan of an FDG-filled line
source inside a scatter phantom consisting of polyethylene. The rows of the
measured sinogram are centered at their maxima and the sum of all rows is
calculated. In the resulting radial profile of the phantom scan, the NEMA
standard specifies a signal window of 7\,mm around the maximum. All event
counts outside this signal window are regarded as either scatter or randoms.
It is assumed that the sum of scatter and random event counts is at the same
level inside the signal window as on the edges of the signal window. The sum of
random and scatter event counts is denoted as $C_{r+s}$, and the sum of all event
counts are denoted as the total event count $C_{\text{TOT}}$.

For systems without intrinsic radioactivity, the scatter fraction is supposed to
be determined by assuming that the contribution of the random rate to the combined
scatter and random counts $C_{r+s}$ is negligible for measurements at a low
activity. Then, the random rate is determined from the total event rate
$R_{\text{TOT}}$ and true event rate $R_t$.

For systems with intrinsic radioactivity the sum of random and scatter event
counts also includes the random event counts produced by the intrinsic
radioactivity and this contribution of the intrinsic random rate cannot be
neglected at low measured activities. The NEMA standard therefore states:
"For systems employing
detector material containing intrinsic radioactivity, the scatter fraction
shall be evaluated by first evaluating the scattered event counting rate (see
section 4.4.5 below)."
\cite[p. 13]{national_electrical_manufacturers_association_nema_nema_2008}
However, section 4.4.5 gives the following formula for the scattered event
counting rate $R_s$, which already includes the random rate $R_r$
\cite[p. 14]{national_electrical_manufacturers_association_nema_nema_2008}
\begin{equation}
	\label{eq:R_s}
	R_s = R_{\text{TOT}} - R_t - R_r - R_{\text{int}}
\end{equation}
The formula for the random rate is given above, in section 4.4.4 and it
includes the scatter fraction $SF$
\begin{equation}
	\label{eq:R_r}
	R^{\text{NEMA}}_r = R_{\text{TOT}} - \left(\frac{R_t}{1 - SF}\right)
\end{equation}
The scatter fraction $SF$, which is defined in the mentioned section 4.4.5, in turn
includes the scattered count rate
\begin{equation}
	\label{eq:SF}
	SF = \frac{R_s}{R_t + R_s}
\end{equation}

Therefore, the definition of scattered rate $R_s$, scatter fraction $SF$ and
random rate $R_r$ are circular when reading the NEMA standard verbatim.

Next, we show that the definition is not only circular but also not
satisfiable. We insert the definition of $SF$ (i.e equation~\ref{eq:SF}) into
the definition of $R_r$ (i.e. equation~\ref{eq:R_r}:
\begin{eqnarray*}
	R^{\text{NEMA}}_r &= R_{\text{TOT}} - \left(\frac{R_t}{1 - \frac{R_s}{R_t
	+ R_s}}\right) \\
	&= R_{\text{TOT}} - \left(\frac{R_t}{\frac{R_t + R_s - R_s}{R_t
	+ R_s}}\right) \\
	&= R_{\text{TOT}} - R_t - R_s
\end{eqnarray*}
This is inserted into the definition of $R_s$ (i.e. equation~\ref{eq:R_s}):
\begin{eqnarray*}
	R_s &= R_{\text{TOT}} - R_t - \left(R_{\text{TOT}} - R_t - R_s\right) -
	R_{\text{int}} \\
	&= R_s - R_{\text{int}} \qquad \mbox{\lightning\;for}\;
	R_{\text{int}} \neq 0
\end{eqnarray*}

This is a contradiction, because by definition it is true that $R_{\text{int}}
\neq 0$, since the standard specifies these definitions of $R_r$ and $R_s$ for scanners with intrinsic radioactivity.

We can speculate on the intended meaning of the NEMA standard's definitions.
One sensible and probably originally intended modification to the definitions
would be to neglect the influence of the random rate $R_r$ (i.e. assume $R_r =
0$) in equation~\ref{eq:R_s} for measurements at low activities to determine
$R_s$ and $SF$. We can then assume that $SF$ is
approximately constant with increasing activity and use $SF$ 
determined at a low activity to calculate the random rates $R_r$ and scatter
rates $R_s$ at higher activities.

This definition looks as if one could re-evaluate equations~\ref{eq:R_s},
\ref{eq:SF} and \ref{eq:R_s} iteratively, starting with the
assumption that $R_r = 0$ at low activities, to iteratively obtain a more
accurate estimates of $R_r$, $R_s$ and $SF$.
However, this iteration diverges which is another indicator that the NEMA
standard's definitions are not satisfiable. Nevertheless, this approach allows
to determine estimates of $R_r$, $R_s$ and $SF$ if we stop after one
iteration, i.e. we use the $SF$  determined at a low activity to
calculate $R_r$ and $R_r$ and then re-evaluate $SF$ at higher activities.

The NEMA standard specifies the following lower activity threshold: "For
scanner employing, radioactive scintillator material, measurements shall be
performed until the single event rate is equal to twice intrinsic single event
rate" \cite[p. 11]{national_electrical_manufacturers_association_nema_nema_2008}.
Our scanner has an intrinsic single event rate of 80\,kcps and we reach a
single event rate of 160\,kcps at 430\,kBq. Thus, we use this activity to
determine the scatter rate $R_s$ using equation~\ref{eq:R_s} while neglecting
the random rate. This scatter rate is then used with equation~\ref{eq:SF} to
determine the scatter fraction $SF$. This scatter fraction is assumed to be
constant with varying activity and we use this with equation~\ref{eq:R_r} to
determine the random rates $R_r$ at different activities. With these random
rates we can evaluate equation~\ref{eq:R_s} and~\ref{eq:SF} again to determine the scatter
rates and fractions at higher activities without neglecting the random rates.

Alternatively, the NEMA standard allows the usage of a randoms estimate supplied
directly by the scanner. Such estimates usually use one of two techniques: one
using a delayed coincidence window (DCW) and one based on the singles rates
\cite{hoffman_quantitation_1981}. The singles rate (SR) method infers the
randoms rate $R_{ij}$ between to detector element $ij$ from the single rates
$S_i$ and $S_j$ using the formula
\begin{equation}
	\label{eq:SR}
	R^{\text{SR}}_{ij} = 2 \tau S_i S_j
\end{equation}
with the time coincidence window $\tau$. However, this method systematically
overestimates the random randoms rate \cite{rafecas_estimating_2007,
oliver_improving_2010}. Oliver et al. \cite{oliver_modelling_2016} proposed an improved
method "Singles Prompt" (SP) which includes corrections based on the
coincidence rate (or prompt rate) $P_i$ to account for the
contribution of true coincidences and pile-up events:
\begin{equation}
	\label{eq:SP}
	R^{\text{SP}}_{ij} = \frac{2 \tau e^{- (\lambda + S) \tau}}
							  {(1 - 2 \lambda \tau)^2}
					  (S_i - e^{(\lambda + S) \tau} P_i)(S_j - e^{(\lambda + S)\tau} P_j),
\end{equation}
where $\lambda$ is the solution of the equation
\begin{equation}
	2 \tau \lambda^2 - \lambda + S - P\,e^{(\lambda + S) \tau} = 0.
\end{equation}

We have implemented these methods with the Hyperion II$^\text{D}$ scanner and
can compare them empirically with the modified method the NEMA standard suggests.
The NEMA standard specifies a cylindrical signal window of 8\,mm around the
phantom (i.e. a total diameter of 41\,mm) in sinogram space. We applied an
equivalent cylindrical signal window, i.e. we only determined the
random rate for the pairs of detector elements whose line of responses
intersect with the cylindrical signal window.

\begin{figure}[tbh]
	\centering
	\includegraphics[width=85mm]{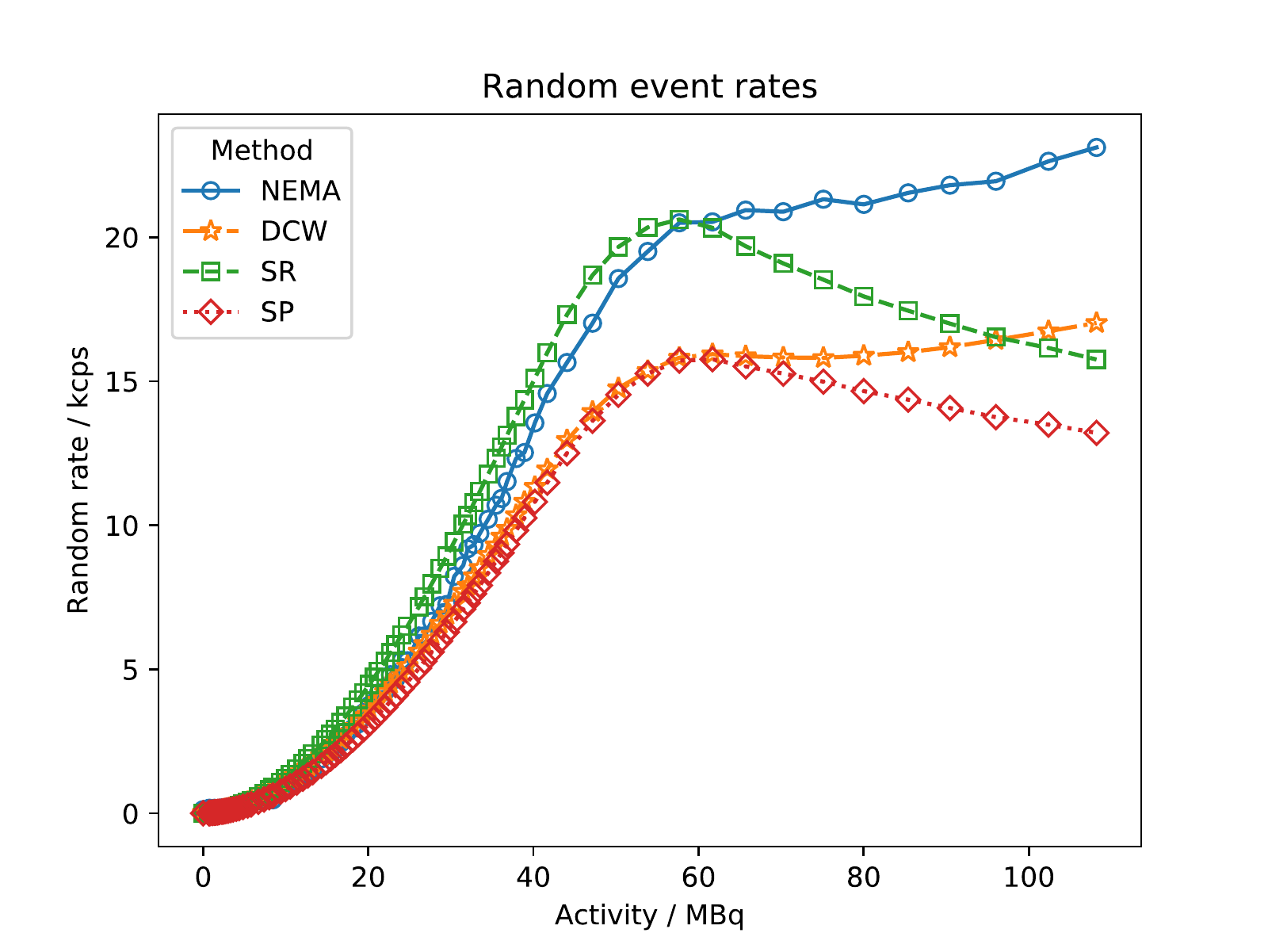}
	\caption{Comparison of different methods for the determination of random
	event rates: NEMA means a method based on the NEMA standard using
	equation~\ref{eq:R_s}, DCW uses a delayed coincidence window,
	SR is based on the singles rate using
	equation~\ref{eq:SR}, and SP incorporates additional corrections using
	equation~\ref{eq:SP}.}
	\label{fig:randoms_comparison}
\end{figure}

Figure~\ref{fig:randoms_comparison} shows the total randoms rates as a function
of activity inside the scanner for the four different methods: NEMA, DCW, SR,
and SP.

As expected, the SR estimate is larger than the SP estimate: $R^{\text{SR}}
\geq R^{\text{SP}}$. The DCW estimate is similar to the SP estimate, and the
adapted NEMA estimate is similar to the SR estimate, which
which is known to be the less precise than DCW and SP
\cite{oliver_modelling_2016}.

Oliver et al. \cite{oliver_modelling_2016}
showed that random estimates $R^{\text{DW}}$ using a delayed coincidence window
(DW) are larger or equal to the SP estimates: $R^{\text{SR}} \geq R^{\text{DW}}
\geq R^{\text{SP}}$.
There are many publications investigating the correctness
of these methods, providing evidence from theory, simulations and measurements.
For the NEMA method, on the other hand, there doesn't exist a single
publication investigating the correctness of the method, to our knowledge.
Additionally, if the NEMA standard is taken verbatim, the described random
estimation method is
not implementable and the described estimation method
has to be adapted as described here, instead. Alternatively, the standard
also allows the use of any randoms estimation method instead. Altogether, the
current standard does not provide objective comparability of measured random rates
of different scanners.

All of these points apply also to the scatter rate $R_S$ defined in
equation~\ref{eq:R_s} and the noise-equivalent count rate, as the definitions
of these observables depend on the randoms rate. Therefore, objective
comparison of these three performance observables between different scanners
based on the NEMA standard is problematic.

\section{Sensitivity}
We think the NEMA standard's protocol for the evaluation of the sensitivity is
unclear. Section 5.3 of the NEMA standard specifies to axially step a point
source through the scanner. Further, section 5.3.4
implies that a different scan for each source position should be acquired. In
section 5.4, all of the data analysis is specified for single sinogram
slices $i$. For instance, the sensitivity is defined as
\begin{equation}
	\label{eq:S_i}
	S_i = \frac{R_i - R_{B,i}}{A_{\text{cal}}}
\end{equation}
with the counting rate $R_i$ and the background rate $R_{B,i}$ of sinogram
slice $i$. However, the NEMA standard only ever references sinogram slices and never
different measurements. We have one measurement per source position and each of
these measurements has many sinogram slices. In other words, there are many
measurements for each axial sinogram slice. Whenever the NEMA standard refers
to sinogram slices, it remains unclear which measurement to consider. One
possible intention could be to calculate the sum of all measurements; however,
this is never explicitly stated. This would effectively create a sensitivity
measurement with a virtual line source of activity $n \cdot A$, where $n$ is
the number of measurements. Such a line source would be similar to the source
distribution specified in the sensitivity protocol in the clinical NEMA NU
2-2012 standard. However, the sensitivity $S_i$ is defined by the activity
$A_{\text{cal}}$ in equation~\ref{eq:S_i}, not a virtual activity $n\,A$ of the
combined measurements. Unfortunately, the NEMA standard does not define
$A_{\text{cal}}$ in this equation, the only definition of $A_{\text{cal}}$ is
in section~1.2 as "activity at time $T_{\text{cal}}$". In conclusion, if this
interpretation were the intention of the NEMA standard, multiple required
instructions would be missing.

Another possible interpretation could be to take the slice $i$ of the
measurement where the point source is located at the center position of the
slice. However, this interpretation is not consistent with the formulas given
for the total system sensitivity
\begin{equation}
	S_{\text{tot}} = \sum_{\text{all}\,i} S_i,
\end{equation}
which lack a normalization for the total number of slices. With a normalization
with the total number of slices, this would effectively be an additional axial
signal window around the point source, However, the size of this axial signal
window would depend on the scanner's slice thickness, giving an unfair
disadvantage to high-resolution scanners. For instance, with a slice thickness
of 1\,mm, this axial signal window would cut into the point source.
Additionally, this interpretation would not be realistic in the context of
real-world applications, where the sensitivity is supposed to be an indicator of how
many true coincidences one can expect for a given activity inside the scanner's
FoV.

In summary, the NEMA standard does not include any instructions on how to analyze
the data of the multiple measurements it instructs to take. It only defines the
sensitivity of sinogram slices without specifying the relationship of the
sinogram slices and measurements with different source positions.

One consistent alternative definition of sensitivity could simply sum all
sinogram slices and then divide the total coincidence counts by the acquisition
time and activity for each measurement (i.e. source position). The sensitivity
profile would consequently be defined as this total sensitivity as a
function of the source position. To calculate the mouse- and rat-equivalent
sensitivities, one would average this sensitivity profile inside the central
7\,cm or 15\,cm. Because the NEMA standard specifies a transversal signal
window with a width of 20\,mm in sinogram space, it would be consistent to
apply the same signal window around the point source in axial direction.

The ambiguity of the NEMA standard leads to unclear and incomparable results in
performance publication based on NEMA. Most publications seem to more or less
ignore the NEMA standard and simply measure the sensitivity for different
source positions as the total sum of all counts in this measurement. However,
the exact methods and definitions used stay usually unclear, impeding an
objective comparison of different sensitivity

For instance,
Prasad et al. \cite{prasad_nema_2011} seem to follow the formulas given by NEMA quite
closely, without clearly specifying how the data of the different measurements at
different source measurement is used in the data analysis. The reported
sensitivity profile has data points above 1\,cps/Bq, i.e. an impossible
sensitivity larger than 100\% for the central slices. They claim a total
absolute sensitivity of 12.74\%, which is implausibly large compared to the expected
geometric sensitivity of 12.9\%, We calculated this ideal geometric sensitivity
using their scanner's diameter, axial length and crystal
thicknesses with the simple geometric model
explained in \cite{hallen_pet_2018}. The usual ratio between measured
peak sensitivity and geometric sensitivity is between 0.3 and 0.5
\cite{hallen_pet_2018}.

Other publications seem to more or less ignore the NEMA standard and simply
measure the sensitivity for different source positions as the total sum of all
counts in this measurement. However, basically all publications leave their
exact method unclear, preventing an objective comparison of different
sensitivity results.

\section{Image Quality, Accuracy of Attenuation, and Scatter Corrections}
The NEMA standard defines several observables for quantitative analysis of the image quality
phantom. The uniformity is defined as the relative standard deviation of all
voxels in a large cylindrical volume of interest over the uniform region in the
image quality phantom.  For determination of the recovery coefficients, the
image slices along the central 10\,mm of the hot rods are averaged. Then, the
recovery coefficients are defined as the maximum values in a circular region of
interest around the hot rods with different diameters, divided by the mean
activity in the volume of interest over the uniform region. The issue with this
definition is that the recovery coefficients are correlated with the
uniformity: The maximum value of a randomly distributed sample increases with
variance, even if the mean value of the distribution is constant. Thus,
this definition of the recovery coefficients does not measure the mean recovery
in the hot rods, but measures a combination of recovery and variance. With a
high variance and a good recovery the recovery coefficients can even reach values
larger than 1.

\begin{figure}[tbh]
	\centering
	\includegraphics[width=85mm]{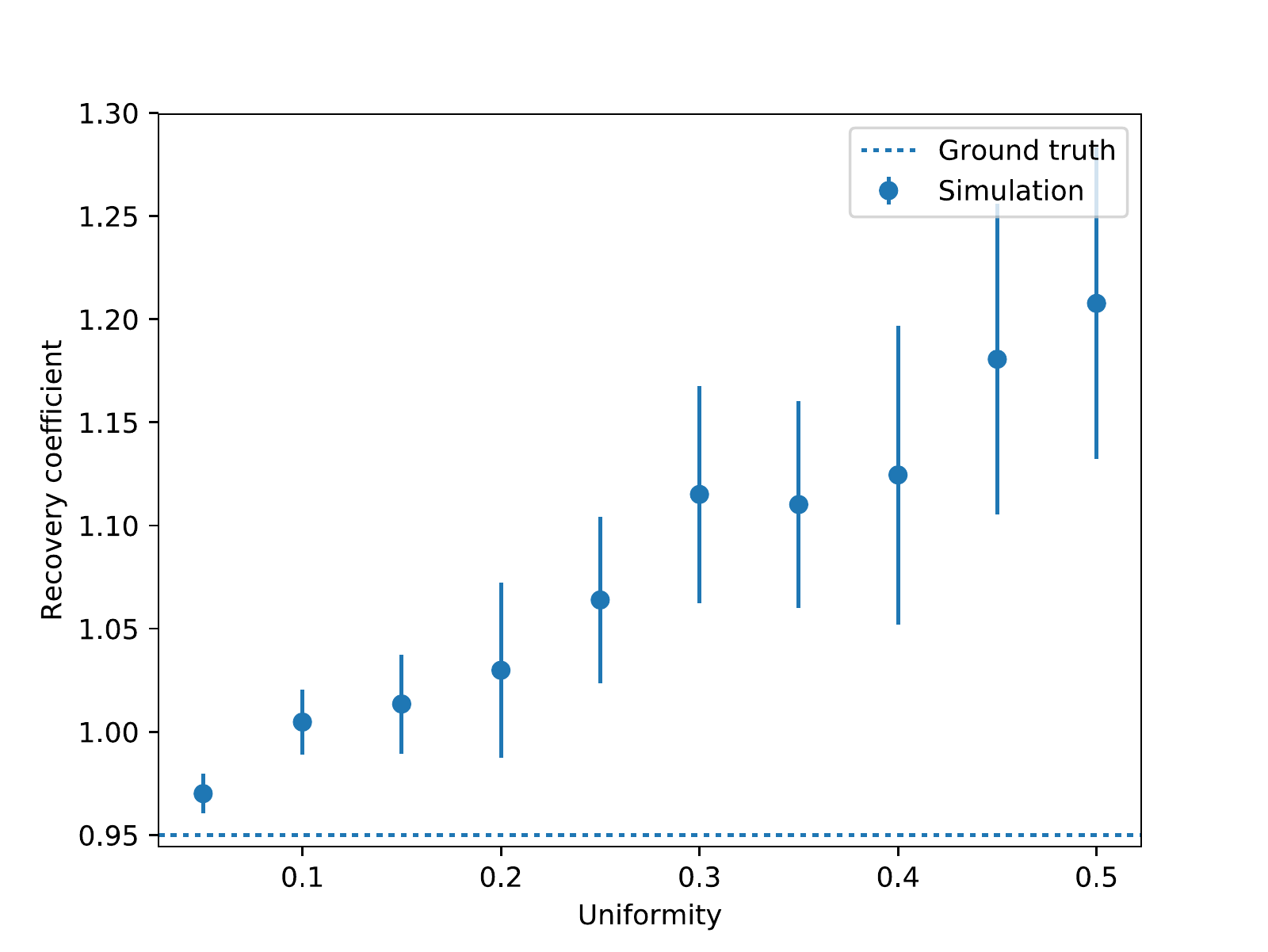}
	\caption{Simulated recovery coefficient of the 5-mm rod as a function of
	the uniformity. The ground truth for the recovery was 0.95. The simulated
recovery coefficients are always larger than the ground truth and increase with
increasing variance (i.e. larger uniformity values).}
	\label{fig:rc_mc}
\end{figure}

We can demonstrate this behavior in a simple Monte Carlo simulation, where we
assume that the reconstructed activity in a voxel follows a normal
distribution with
the standard deviation given by the uniformity. The simulated geometry is the
NEMA image quality phantom. Figure~\ref{fig:rc_mc} shows the simulated recovery
coefficients of the 5-mm rod as a function of the uniformity. The ground truth
for the recovery coefficient for the activity in the rod was 0.95. The data
analysis follows the NEMA standard, i.e. the recovery coefficient is defined by
the maximum activity in the region of interest. The drawn errors are calculated
from the errors on the mean of the averaged pixels in the region of interest.
The simulation demonstrates that the recovery coefficient is always
overestimated compared to the ground truth and increases with increasing
variance (i.e. larger uniformity values).

\begin{figure}[tbh]
	\centering
	\includegraphics[width=85mm]{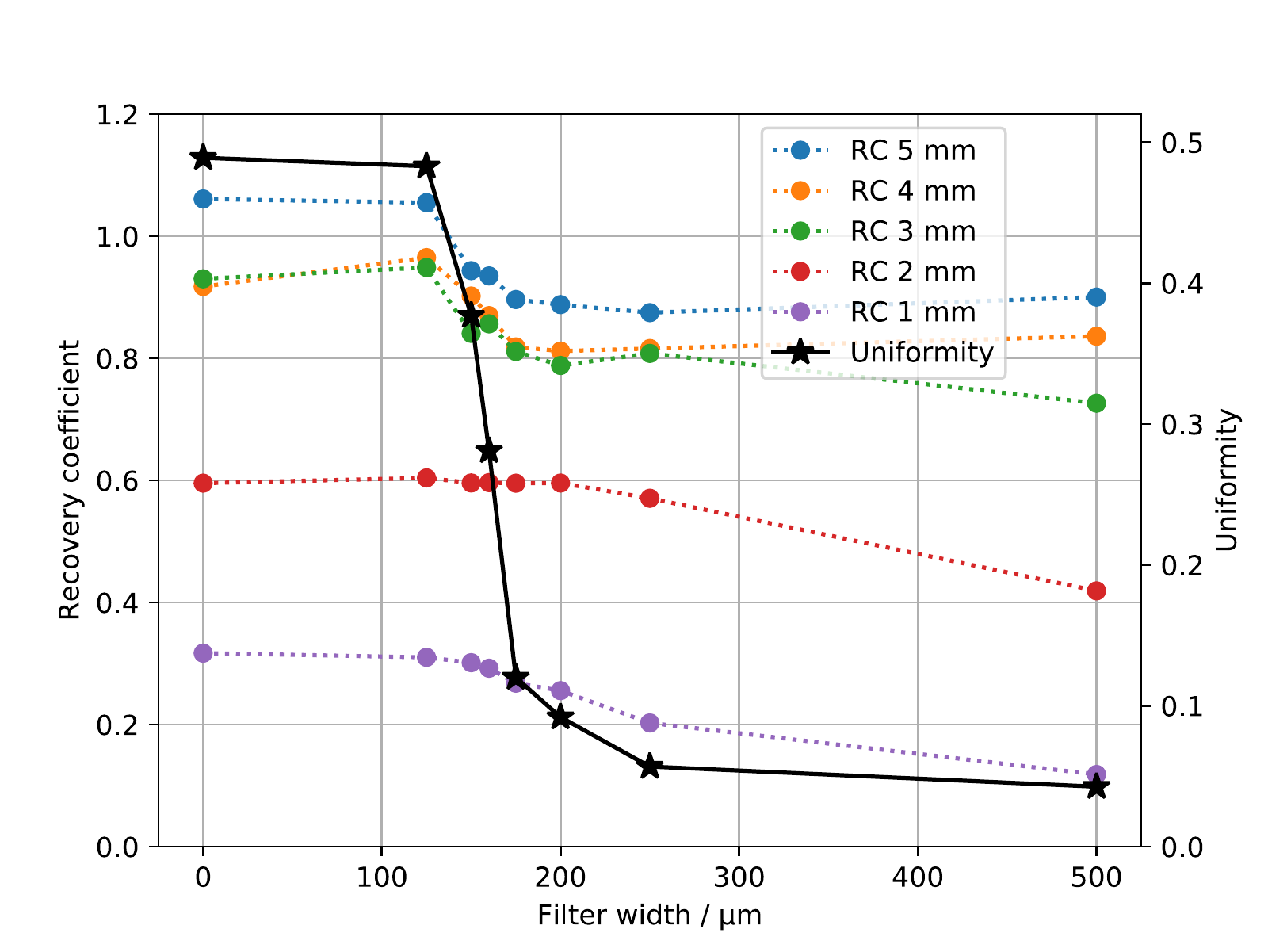}
	\caption{Measured uniformity and recovery coefficients as a function of
	filter width used during reconstruction. A larger filter reduces variance
and therefore increases uniformity (i.e. decreasing relative standard
deviation). The recovery coefficients decrease at the same time, so overall
image quality performance is a trade-off between uniformity and recovery.}
	\label{fig:uniformity}
\end{figure}

Thus, the NEMA standard's definition of the recovery coefficients hampers an
easy comparison of different scanner's recovery performance, because the
recovery and uniformity must be compared at the same time. In other words, the
same scanner can achieve different recovery performance at different uniformity
points. The user can influence the uniformity with parameters such as the
amount of filtering during reconstruction. Figure~\ref{fig:uniformity} shows
how the measured uniformity and recovery coefficients changes with different
widths of a Gaussian kernel used during reconstruction for a scan of the image
quality phantom. We used the maximum likelihood expectation maximization
reconstruction described in \cite{salomon_self-normalization_2012}. As predicted by the
Monte Carlo simulation, the recovery coefficients are correlated with the
relative standard deviation in the uniformity region: Both values decrease with
large filter width, i.e. reduced variance in the image. Of course, it is not
unexpected that the recovery decreases with stronger filtering during
reconstruction. However, the observed effect is on top of the expected decrease
in recovery due to filtering. Using the NEMA standard's observables, improving the uniformity
performance will always lead to a loss in observed recovery, regardless whether
the actual true recovery degraded or not. When conducting a NEMA performance
evaluation one has to chose an arbitrary point on the uniformity and recovery
curve resulting in one of many possible results, which are difficult to compare
with the results of other scanners.

As another minor issue, the NEMA standard
derives the standard deviation of the recovery coefficients from the standard
deviations of the line profiles along axial directions and the standard
deviations of the uniform regions using Gaussian error propagation. This is not
the correct standard deviation of the recovery coefficient,
because the standard deviation of the maximum value of a randomly
distributed value is not the standard deviation of the underlying distribution.

Fixing the definition of the recovery coefficients is not trivial. The NEMA
standard probably uses the maximum due to the small diameters of the hot rods.
For the very small rods, very few, if any, voxels lie clearly in the center of
the rods. Alternative definitions using the mean in a volume of interest will
therefore be biased by the smaller reconstructed activity in the border regions
of the rods. However, with today's high-resolution PET scanners, we believe it
would be possible for most scanners to define volume of interest (VoI) inside the hot
rods and then define the recovery coefficients using the mean reconstructed
activity inside the VoI. Even if these VoIs would partially include the border
regions of the rods, it would still at least be a comparable measure of
recovery for every scanner. For the larger rods it should not be any problem to
define VoI which are well inside the hot rods with a sufficient number of
voxels. It is these larger rods where the current definition of recovery
coefficients leads to basically a recovery of 1 or larger for all current
scanners, hindering a differentiation of subtle differences in recovery between
the scanners.

\section{General points}
The NEMA standard does not explicitly mandate
the use of the same settings for each measurement. Most scanners offer a
multitude of settings for measurements and data processing, such as trigger
settings, coincidence and energy window sizes and quality filters for gamma
interactions. The choice of setting parameters requires often a trade-off for different
performance parameters. For example, the sensitivity benefits from wide energy
and coincidence windows and no quality filters, while the image quality and
spatial resolution benefits from narrow windows and strict quality filters. One
could report very misleading performance results by optimizing the settings for
each performance measurement separately, thus achieving performance results
which are unattainable at the same time in real-world applications.

While following the standard, many performance publications based on NEMA do
neither state if they used the same settings for every measurement explicitly
nor report all used settings for each measurement.
For example, Nagy et al. \cite{nagy_performance_2013} use wide energy windows for the
sensitivity and count rates measurements and a narrow energy window for the
measurement of spatial resolution. They do not report any settings for the
image quality measurement.

Another issue is the mandated use of sinograms. The data analysis for every
measurement except the image quality
measurement are described on sinograms. However, most modern scanners store
their data in listmode format and often only implement sinogram support to
conduct the NEMA measurements. To our knowledge, all NEMA NU-8 measurements
published in in the last 5 years had to convert their native  listmode
files to sinograms after the measurements \cite{nagy_performance_2013, ko_evaluation_2015,
omidvari_pet_2017, spinks_quantitative_2014, miyake_performance_2014,
sato_performance_2016, hallen_pet_2018, mackewn_pet_2015,
wong_engineering_2012, prasad_nema_2011, szanda_national_2011,
krishnamoorthy_performance_2018}.
Spinks et al. \cite{spinks_quantitative_2014} even mentions that the calculation of
scatter fractions were omitted due to missing sinogram support, so this
performance evaluation did apparently only use listmode data for the data
analysis. The number
of scintillator crystals or equivalent positioning bins in monolithic
scintillators is usually above 30\,000 in modern small-animal PET systems,
so that full sinograms have a file size of multiple gigabytes even for very
very short measurements. Listmode
files on the other hands are usually much smaller, making sinograms much more
unwieldy.

The data analyses in the NEMA standard could be specified without the use of
sinograms, since most of the specified cuts in the sinograms could be specified as
cylindrical cuts in the scanner's field of view. The standard could still allow the use of
sinograms as one possibility to implement the specified geometric cuts for
backwards compatibility.

\section{Conclusion}
Eleven years after the publication of the NEMA NU-4 standard, we believe it's
time for a revision of the standard. In this work, we have pointed out several
flaws in the standard which should be addressed in the next version.
Additionally, the new technological developments in the last decade would
warrant discussing an updated version in itself. With this publication, we
would like to open this discution.

\section*{Bibliography}
\bibliographystyle{bmc-mathphys} 
\bibliography{zotero.bib}      


\end{document}